\documentclass[runningheads,a4paper]{llncs}

\usepackage{amssymb}
\usepackage{graphicx}

\usepackage{url}
\urldef{\mailsa}\path|{kwawrzyn,wislicki}@fuw.edu.pl|
\newcommand{\keywords}[1]{\par\addvspace\baselineskip
\noindent\keywordname\enspace\ignorespaces#1}

\newcommand{\bdm}{\begin{displaymath}}
\newcommand{\edm}{\end{displaymath}}
\newcommand{\be}{\begin{equation}}
\newcommand{\ee}{\end{equation}}
\newcommand{\ba}{\begin{eqnarray}}
\newcommand{\ea}{\end{eqnarray}}
\newcommand{\efig}{ \end{figure}}

\begin{document}

\mainmatter  

%
%
%
%
%

\title{Grand canonical minority game as a sign predictor}

\titlerunning{GCMG as a sign predictor}

%
%
\author{Karol Wawrzyniak%
\and Wojciech Wi\'slicki}
\authorrunning{K.Wawrzyniak \and W.Wi\'slicki}

\institute{National Centre for Nuclear Research\\
Ho\.za 69, 00-681 Warszawa, Poland\\
\mailsa\\
\url{http://agf.statsolutions.eu}}

\toctitle{Lecture Notes in Computer Science}
\tocauthor{Authors' Instructions}
\maketitle

\begin{abstract}
In this paper the extended model of Minority game (MG), incorporating variable number of agents and
therefore called Grand Canonical, is used for prediction. We proved that the best MG-based
predictor is constituted by a tremendously degenerated system, when only one agent is involved. The
prediction is the most efficient if the agent is equipped with all strategies from the Full
Strategy Space. Each of these filters is evaluated and, in each step, the best one
is chosen. Despite the casual simplicity of the method its usefulness is invaluable in many cases
including real problems. The significant power of the method lies in its ability to fast adaptation
if $\lambda$-GCMG modification is used. The success rate of prediction is sensitive to the properly
set memory length. We considered the feasibility of prediction for the Minority and Majority games.
These two games are driven by different dynamics when self-generated time series are considered.
Both dynamics tend to be the same when a feedback effect is removed and an exogenous signal is
applied.
\end{abstract}

\section{Introduction}\label{ch:GCMG}
The minority decision is defined as a function of a self-generated signal called aggregate
attendance or aggregate demand~\cite{challet00PhysicaA276,moro00minorityIntroductoryGuide}. In the
standard Minority Game, the sequence of minority decisions constitutes the basis for individuals'
actions. Following an economics terminology, the series of minority decisions formed inside a
model, is called \emph{endogenous}. Hence, the MG mechanism is tantamount to predicting a future
value using past values of the same, endogenous, time series. The mentioned feedback effect exists
at the level of the population but not at the level of single agent. That is, although individuals'
decisions directly influence the aggregate demand, the agents themselves do not possess any
mechanisms to account their contributions to the aggregate variable. Hence, individuals are unable to recognize whether the signal is self-generated or a fake
one i.e. taken from the outside of the model. In opposition to the term endogenous, the series of
fake histories is called \emph{exogenous} if it affects a model, but is not affected by it. Hence,
instead of showing to agents true i.e. self-generated histories one can generate them randomly.
Such game is commonly known in the literature as MG with fake histories \cite{coolen05OxfordPress}.
The genuine and modified MG is described in Sec.~\ref{sec:definition}.

As presented in Ref.~\cite{cavagna99PhysRevE59}, the MG can be potentially used as the predictor of
any exogenous (fake) series, provided that dependencies in the signal reflect the patterns built in
strategies\cite{johnson01PhysicaA299,gou05IEEE,gou06ChinesePhys}.
The details of the current state of the art are presented further in section
\ref{sec:pr:relations}.

In section \ref{sec:model} we presented the model and its configuration. Then, in section \ref{sec:optimization}, we verified the quality of the
predictor using the time series generated by the well understood, autoregressive stochastic
process. After analyzing our
numerical results, we provided a methodology for tuning the parameters. Intriguingly, the best
results are achieved if the game is degenerated to only one single agent equipped with all
strategies from the whole strategy space. This new discovery seems to stay in contradiction to
commonly used optimization techniques~\cite{johnson01PhysicaA299,gou05IEEE,gou06ChinesePhys}, where authors try to find a set of
parameters for which the statistical properties of exogenous and predicted time series are mutually
close.

Additionally, in section \ref{sec:optimization} we presented some new insights which allow us to
improve the model. For example, it was proved that if the exogenous signal is exploited, then there
is no qualitative difference between minority and majority game. We also introduced a modification,
the so called $\lambda$-GCMG, that is well suited for quasti-stationary signals. In all of
experiments, where the autoregressive process was involved, we compared the MG results with those
achieved by the best theoretical predictor found for the analyzed process.

Finally, in the section \ref{sec:optimization}, the properly tuned MG model was applied as a
forecaster of assets prices on financial markets. For some intraday data the achieved success rate
of one-step prediction is around 70\% what significantly exceeds the random case.
\section{The Formal Definition of the Minority Game}\label{sec:definition}
At each time step $t$, the $n$th agent out of $N$ $(n=1,\ldots,N)$ takes an action
$a_{\alpha_n}(t)$ according to some strategy $\alpha_n(t)$. The action $a_{\alpha_n}(t)$ takes
either of two values: $-1$ or $+1$. An aggregated demand is defined
\begin{eqnarray}
A(t)=\sum_{n=1}^{N}a_{\alpha_n^\prime}(t), \label{eq:A}
\end{eqnarray}
where $\alpha_n^\prime$ refers to the action according to the best strategy, as defined in
eq.~(\ref{eq:argmax}) below. Such defined $A(t)$ is the difference between numbers of agents who
choose the $+1$ and $-1$ actions. Agents do not know each other's actions but $A(t)$ is known to
all agents. The minority action $a^\ast(t)$ is determined from $A(t)$
\begin{eqnarray}
a^\ast(t)=- \mbox{sgn} A(t). \label{eq:aMinority}
\end{eqnarray}
Each agent's memory is limited to $m$ most recent winning, i.e. minority, decisions. Each agent has
the same number $S\ge 2$ of devices, called strategies, used to predict the next minority action
$a^\ast(t+1)$. The $s$th strategy of the $n$th agent, $\alpha_n^s$ $(s=1,\ldots,S)$, is a function
mapping the sequence $\mu$ of the last $m$ winning decisions to this agent's action
$a_{\alpha_n^s}$. Since there is $P=2^m$ possible realizations of $\mu$, there is $2^P$ possible
strategies. At the beginning of the game each agent randomly draws $S$ strategies, according to a
given distribution function $\rho(n):n\rightarrow \Delta_n$, where $\Delta_n$ is a set consisting
of $S$ strategies for the $n$th agent.

Each strategy $\alpha_n^s$, belonging to any of sets $\Delta_n$, is given a real-valued function
$U_{\alpha_n^s}$ which quantifies the utility of the strategy: the more preferable strategy, the
higher utility it has. Strategies with higher utilities are more likely chosen by agents.

There are various choice policies. In the popular {\it greedy policy} each agent selects the
strategy of the highest utility
\begin{eqnarray}
\alpha_n^\prime(t)=\arg \max_{s:\,\alpha_n^s \in \Delta_n} U_{\alpha_n^s}(t). \label{eq:argmax}
\end{eqnarray}
If there are two or more strategies with the highest utility then one of them is chosen randomly.
Each strategy $\alpha_n^s$
is given the {\it payoff} depending on its action $a_{\alpha_n^s}$
\begin{eqnarray}
\Phi_{\alpha_n^s}(t)=-a_{\alpha_n^s}(t)\,g[A(t)], \label{eq:R}
\end{eqnarray}
where $g$ is an odd {\it payoff function}, e.g. the steplike $g(x)= \mbox{sgn}(x)$
\cite{challet97PhysicaA246}, proportional $g(x)=x$ or scaled proportional  $g(x)=x/N$. The learning
process corresponds to updating the utility for each strategy
\begin{eqnarray}
U_{\alpha_n^s}(t+1)=U_{\alpha_n^s}(t)+\Phi_{\alpha_n^s}(t), \label{eq:U}
\end{eqnarray}
such that every agent knows how good its strategies are.

The presented definition is related to genuine MG~\cite{challet00PhysicaA276}. If game is used as
the predictor then the feedback effect is destroyed and $\mu$ is e.g. generated randomly.
\section{Relation to other models}\label{sec:pr:relations}
As it is know from other
works~\cite{challet00PhysicaA276,challet99PhysRevE60,savit99PhysRevLetters82} the standard MG
exhibits an intriguing phenomenological feature: a non-monotonic variation of the volatility when
the control parameter is varied. There are two mechanisms potentially responsible for it: the
feedback effect and the quenched disorder~\cite{challet00PhysicaA276}. The incorporated feedback
effect couples input and output signals in such a way that a minority decision at time $t$
constitutes the basis for future agents' decisions. The quenched disorder is related to an initial,
random realization of agents' strategies in their strategy space. In theoretical papers \cite{cavagna99PhysRevE59,challet00PhysRevE62,lee01PhysRevE65} it is
discussed how the feedback mechanism affects observed behavior of MGs. For us it is important that the lack of the feedback does not influence the population's predictive power which is exclusively driven by the quenched disorder.

Some other authors applied the model to the exogenous, real data, assuming an existence of patterns
in these data and using MG as a predictor of its future
value~\cite{johnson01PhysicaA299,jefferies01EurPhysJB20,gou05IEEE,gou06ChinesePhys,chen08PhysA387,krause09IDEAL}.
Although the MG-based predictor is able to forecast any time series assuming that the length of
patterns suits agents' strategies, the commonly used exogenous time series are those related to
asset prices. The idea of feeding the game with a real signal was first
implemented by Johnson \emph{et al.} in Refs.~\cite{johnson01PhysicaA299,jefferies01EurPhysJB20}.
The authors performed an experiment where the time series of hourly \mbox{Dollar \$/Yen \yen}\quad
exchange rate was examined. The achieved results show the 54\% success rate of the next movement
prediction. This level is quite significant and suggests that the model performs better than
random. Although the results presented in Ref.~\cite{johnson01PhysicaA299} are interesting and
inspiring, there is nearly no details about the conditions of the experiment. Such parameters like
$m$, $S$, $N$ are not revealed, making the results unreproducible.

The prediction method used in Ref.~\cite{jefferies01EurPhysJB20} was further developed by
others~\cite{gou05IEEE,gou06ChinesePhys,chen08PhysA387}, and applied to daily data of the Shanghai
Index. The similar model, where agents have various lengths of memory, was introduced in
Ref.~\cite{krause09IDEAL}. Above methods of optimization are based on comparison between two distributions of signals,
i.e. the exogenous and predicted one. If the distributions are mutually close to each other, the
model is considered as a well fitted to the object. As we presented in section
\ref{sec:optimization}, this technique, although interesting, does not assure that the success rate
of the one-step prediction is maximized.

\section{The model}\label{sec:model}
Here, we present the details about our implemented predictor and its configuration. We used, the
Grand Canonical extension of the MG in all our simulations.
\subsection{Grand canonical extension}\label{sec:GCMG} In the standard MG all agents have to play at
each time step $t$, even if all of their strategies are unprofitable. Looking for analogies to
financial markets we see that in real life investors behave differently. If for some of them
trading is not profitable, they withdraw from the market. Hence, at given time, two groups of
agents can be found on the real market: (i) active - actually engaged in the game, (ii) passive - observing the game and waiting for the proper moment to enter.
Formally, staying apart from the market is realized by \emph{zero strategy}. This additional
strategy, marked as $\alpha^0_i$, maps all $\mu$ to the $a_i = 0$ and does not influence the
aggregate demand $A$. We assume a
constant risk-free interest rate as being equal to $U_i^{\alpha^0}(t) = U_i^{\alpha^0} = 0$. During
the game, each agent $i$ monitors average profits $U_i^{\alpha}(t)$ of each of his/her strategies
$\alpha_i$. If $U_i^{\alpha^0}(t)$ is higher than the utility of any other strategies, than the
strategy $\alpha^0$ is used and the agent stays beyond the market.
\subsection{Configuration}\label{sec:configuration}
\begin{figure}[!htb]
\begin{center}
\includegraphics[width=0.65\linewidth]{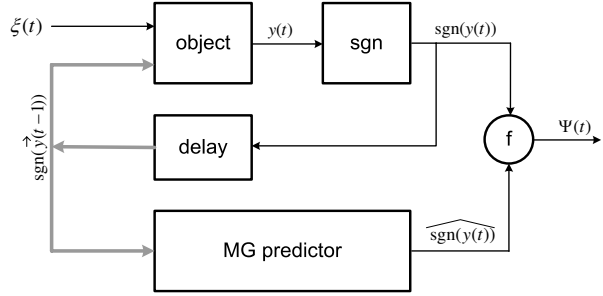}
\caption{\em Block scheme presents the MG in the configuration of prediction.} \label{fig:blockConfiguration}
\end{center}
\end{figure}

Technically, the predictor works according to the diagram presented in Fig.
\ref{fig:blockConfiguration}. The object that we suppose to model is treated as a black-box
stimulated by (i) the vector of previously generated signs of samples
$\mbox{sgn}\big(\overrightarrow{y}(t-1)\big) = [\mbox{sgn}\big(y(t-1)\big) \ldots
\mbox{sgn}\big(y(t-n)\big)$], where $n
> 1$, and (ii) the external information $\xi(t)$. We assume that samples of $\xi$ are
Independent and Identically Distributed (IID). The model is supposed to retrieve dependencies
between the past and future values of $\mbox{sgn}(y(t))$.
The delay block introduces a one-step delay to its every input sample $y(t)$ and forms the vector
$\vec{y}(t-n)$ of $n>1$ past samples. The MG model predicts the next sign of sample exploiting the
information included in the signs of previous movements. The block $f$ compares the predicted
signal with a real output of the object and calculates the \emph{correctness} $\Psi(t)$. The
correctness $\Psi(t)$ represents the average success rate of prediction and is calculated as a
percentage of properly predicted signs of $y$, provided that all samples up to time $t$ are
considered:
\be \Psi(t) = \frac{1}{t} \sum_{n=1}^t \delta( \mbox{sgn}(y(n)), \widehat{\mbox{sgn}(y(n))} ) \ee
where $\delta$ stands for Kronecker symbol.

Two types of objects were analyzed: the autoregressive stochastic processes and the time series of
real prices of shares. The former is mainly used to demonstrate interesting properties of the model
and to learn how to tune its parameters. The latter one is used as an example of practical
application of the predictor. In the case of the autoregressive stochastic process, assuming that
the definition of the object is known, the maximal theoretical level of $\Psi$ can be calculated as
follows:
\be \Psi^{MAX}(t) = \frac{1}{t} \sum_{n=1}^t \delta( \mbox{sgn}(y(n)), \mbox{sgn}({\mathbb
E}[y(n)]) ). \label{eq:corrMax}\ee
The expected value ${\mathbb E}[y(n)]$ is calculated recursively for each $n$ according to the
process definition.
\section{Optimization of the parameters}\label{sec:optimization}
Initially, we apply the predictor to the third order autoregressive time series, AR(3) defined as
follows:
\be y(t) = 0.7y(t-1) - 0.5y(t-2) - 0.2y(t-3) + \xi(t), \label{eq:AR3filter}\ee
where $\xi(t)$ is an instance of the standardized gaussian white noise. It is easy to check that
the process is stable\footnote{Taking the Z-transform, the transfer function $H(z)$ can be
calculated. Three poles of $H(z)$ are located within unit circle $|z|< 1$ what indicates that the
analyzed system is stable.}. It was found numerically that the
process is characterized by ${\mathbb E}[\Psi^{MAX}] = 0.77$, and ${\mathbb Var}[\Psi^{MAX}] <
0.01$, where $t=3000$ steps, and the average was taken over ten realizations.
\subsection{Majority \emph{vs} minority game}\label{sec:MajorityGame}
The model extensively used in literature is based on the grand canonical minority game
~\cite{johnson01PhysicaA299,jefferies01EurPhysJB20,gou05IEEE,gou06ChinesePhys,chen08PhysA387}.
However, it was not obvious for us if the minority mechanism is better than the majority one. In
fact we found that, in the case of prediction, both mechanisms are equivalent.

The algorithm of the majority game is very similar to that of the standard minority game. The only
difference is a formula (\ref{eq:R}) which for majority game reads
\be \Phi_{\alpha_n^s}(t)= a_{\alpha_n^s}(t)\,g[A(t)]. \label{eq:Rmod}\ee
We consider two time series: the endogenous and exogenous. Considering first the game with
endogenous time series we find a number of differences between the minority and majority game. In
the minority game no one of strategies is permanently profitable provided that the game is large
enough~\cite{johnson99PhysA269,wawrzyniak10ACS}. Hence, the number of winners and losers changes in
time. On the contrary, in the majority game the number of winners and losers is stable and, on
average, $N(1-\frac{1}{2^S})$ agents are in majority. The reasoning behind is similar to that
presented in Refs.~\cite{wawrzyniak09ACS12No4and5,wawrzyniak09ACSNo6} and utilizes our observation
that the first large oscillation creates a comprehensive difference in utilities of strategies. The
strategies are divided into two categories: the {\it good} with the positive payoff, and {\it bad}
ones with negative. In any time step $t$ there is on average $N(1-\frac{1}{2^S})$ agents with at
least one good strategy and only $N/2^S$ agents with all bad strategies. Since most of agents has
at least one good strategy - they use it. The difference, compared to the minority game, is that
those who are in majority are not interested in changing the choice when they win. Similarly, the
losers cannot change their situation because they do not have even one good strategy. Given this,
the division between these two groups stays stable and the aggregate demand and utilities exhibit
the one-directional trend. It persists in contradiction to the minority game, where the processes
are mean-reverting. The above is true if the payoff is linear and the feedback effect is
incorporated, i.e. the series of decisions is endogenous. The reasoning is slightly modified when
the step-like payoff is applied but also in this case the essential differences between the
minority and majority game remain.

Intriguingly, both games are equivalent, regardless of the payoff, when series of decisions is
exogenous. Originally, in the standard minority game, strategies predicting sign opposite to $y(t)$
are rewarded. Assuming that patterns in the exogenous signal exist, the individuals prefer more
often strategies predicting $\mbox{sgn}(y(t))$ incorrectly i.e. most of them fail with prediction.
The predictor aggregates decisions of individuals and acts in opposition to the majority, and,
predicts correctly. Contrary to the minority game, in the majority game, strategies that correctly
forecast $\mbox{sgn}(y(t))$ are rewarded and most of agents follow strategies more frequently
recognizing patterns. Subsequently, the predictor acts according to action suggesting the majority
and also correctly recognizes patterns. Given this, the majority and minority game should provide
the same quality of prediction. This is confirmed by numerical simulations presented in
Fig.~\ref{fig:correctness} (right). The experiment was performed using ten realizations of the
AR(3) process. The number of strategies per agent $S$ was set to $2$, the memory length $m$ to $3$
and the number of agents $N$ was varied. Two curves show the success rate of prediction related to
the minority nad majority approaches. As it is seen, there are no qualitative differences between
them. Small distortions are due to random choice of strategies at the beginning of all simulations.
\subsection{Tuning of $m$, $N$ and $S$ parameters}
\begin{figure}[h]
    \begin{minipage}[!t]{0.51\linewidth}
        \centering
            \includegraphics[width=1\textwidth]{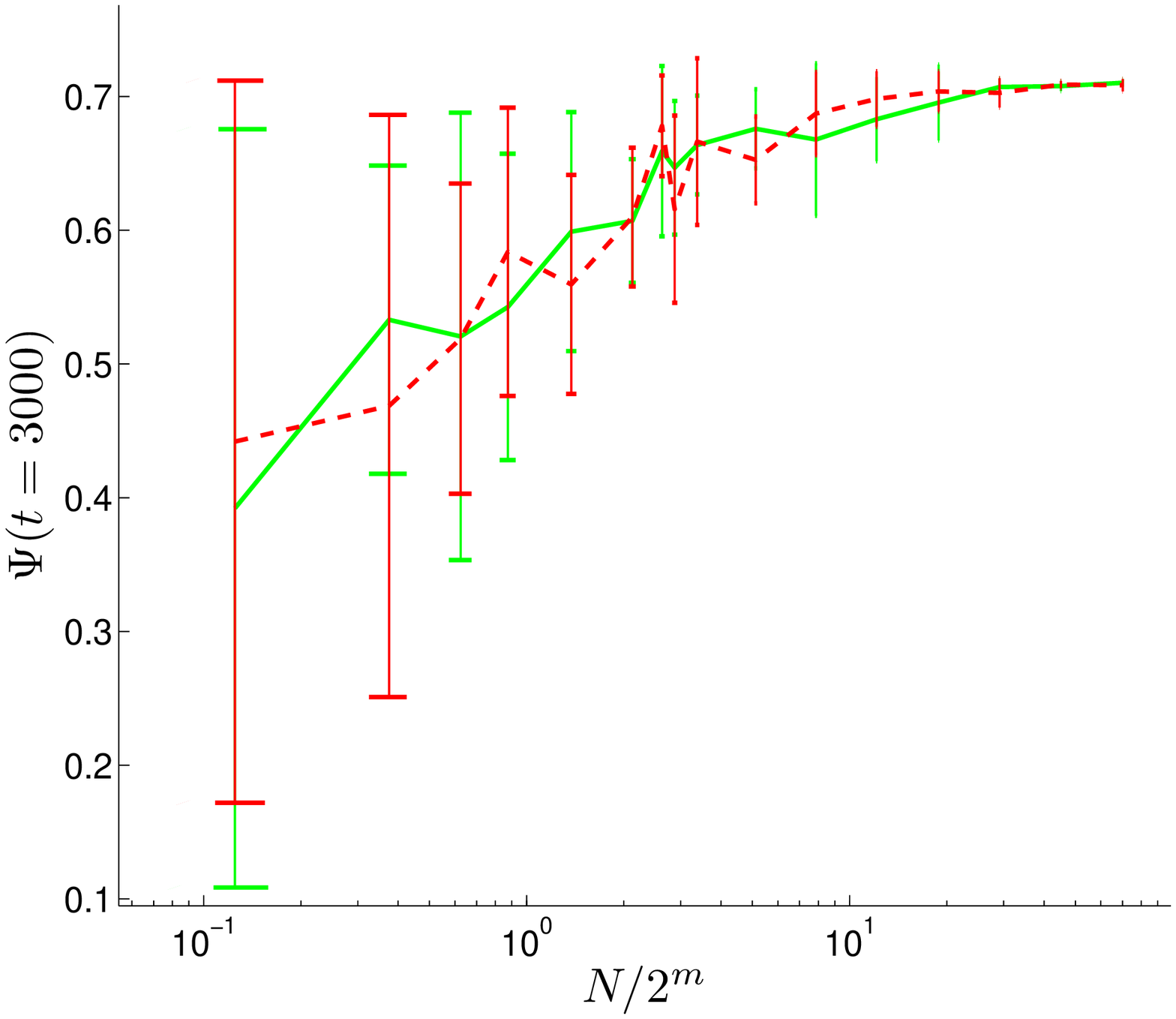}
    \end{minipage}
    \hfill
    \begin{minipage}[!t]{0.51\linewidth}
        \centering
            \includegraphics[width=1\textwidth]{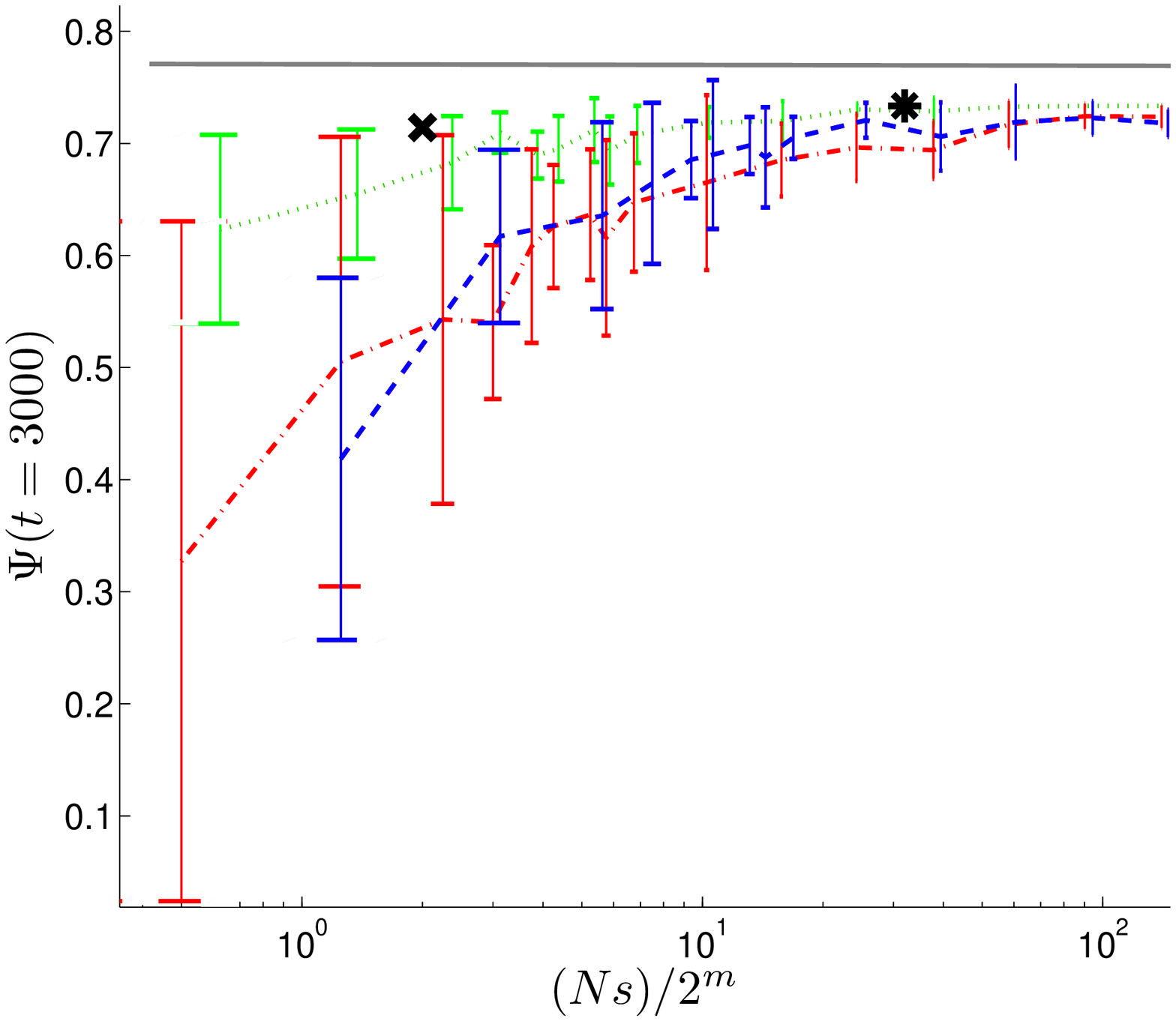}
    \end{minipage}
\caption{\em Left: Comparison of correctness $\Psi$ for the minority and majority games for $S=2$
and variable $N$. Red dashed line corresponds to the minority game and green solid line to the
majority game. Right: Solid grey line corresponds to the theoretical maximal value, green dotted
line corresponds to MG for $N=1$ and various $S$ values, blue dashed line corresponds to MG for
$S=5$ and various $N$ values, red dash-dotted line corresponds to MG for $S=2$ and various $N$
values. The 'x' mark corresponds to MG for $N=1$ and $S=16$ pairwise different strategies from the
Reduced Strategy Space.The star '*' corresponds to MG for $N=1$ and $S=256$ pairwise different
strategies from the Full Strategy Space. Error bars correspond to one standard deviation and curves
are drawn to guide ones eye.} \label{fig:correctness} \efig
Considering three parameters: $m$,$N$ and $S$, at least $m$ requires different optimization techniques than two others.
The optimization of $m$ is strictly related to analysis of time series properties, especially the
analysis of the range of dependencies between values of $Y$. Therefore it depends on the
researcher's knowledge about the object. There are different methods of finding this range.

If the process is explicitly known, $m$ is equal to the order of this process, e.g. $m=3$ for
AR(3). The predictor with $m$ lower than the order cannot work effectively because strategies would
incorrectly recognize patterns. Larger values of $m$ would introduce additional and unnecessary
noise that would degrade the prediction. The latter effect is further illustrated and explained in
the section ~\ref{sec:applicationLambda}.

If the order is unknown but the type of the process is known (e.g. autoregressive one) some
techniques based on the autocorrelation analysis can be applied to detect the order, as presented
in Ref.~\cite{box94TimeSeries}.

The problem is most difficult if the order and the system are unknown, as it is in many real cases
(e.g. prices in financial markets). Still the correlation analysis can suggest the number of past
samples be used. In this section we assume that the order of the examined AR process is explicitly given.

In order to find the proper technique for $N$ and $S$ optimization, let us assume that a certain
pool of strategies of constant size has to be optimally assigned to agents. It means that, under
the constraint $NS=const$, we are looking for the proportion between number of agents $N$ and
number of strategies per agent $S$ that maximizes the correctness $\Psi$. We would also like to
examine if the optimal proportion is sensitive to the constraint's change. Our numerical studies
are presented in Fig. \ref{fig:correctness}. The correctness is presented as a function of $NS$, as
changing the size of the strategies' pool influences the results. Solid grey line corresponds to
the maximal value that is reached by the best possible predictor (in this case linear filter). Blue
dashed line corresponds to MG for $S=5$ and various $N$ values, red dash-dotted line corresponds to
MG for $S=2$ and various $N$ values. These two curves show that the more strategies per agent the
better the correctness $\Psi$, provided that the constraint $NS=const$ is preserved. Following
further this reasoning, the most efficient is obtained for just one agent possessing given pool of
strategies. Indeed, this is confirmed by the green dotted line corresponding to MG for $N=1$ and
various $S$ values. As it is seen, the predictor with such configuration outperforms any other
predictor.

In order to reason out of the presented results, assume there are many agents but only one of them
has the best strategy. Even if this strategy suggests a correct prediction  for itself the
prediction of the whole system can be potentially incorrect, provided there is sufficiently many
agents with bad strategies. Such situation is less probable when the number of strategies per agent
is increased and concurrently the number of agents is decreased. This is impossible if there is
only one agent because then the best strategy is always used. Similarly, if the constraint on $NS$
is shifted towards larger values and number of agents is preserved, then the probability, that more
and more agents have a correct strategy becomes large. Hence, in the limit $NS\rightarrow\infty$,
all agents have the correct strategy and the efficiency of the group is the same as the efficiency
of a single agent equipped with the aggregated pool of strategies.

Considering $\Psi$ as a function of the $S$ value, the larger value of $S$, the better results, as
there is a larger probability that better strategy is drawn by an agent. However, if $S$ is above
some threshold, the pool of strategies is oversampled and many agent's strategies are the same.
Since there is no particular gain if agent has more than one best strategy at his disposal, the
success rate does not increase. Given
this, one can wonder if a random draw is the most efficient way to generate strategies. Indeed it
is not. Only the fully probed strategy space
assures that, if the best strategy exists, then it is used in the game, provided $N=1$. The fringe
benefit is that there is also no redundancy between strategies, what reduces the computational
costs. Hence, the best MG predictor is based on single agent equipped with all pairwise different
strategies, i.e. all strategies covering Full Strategy Space (FSS). However, in cases when $m$ is
large, it can be hardly possible to generate so huge set of strategies. Therefore, for higher $m$ it seems reasonable to use all strategies from
Reduced Strategy Space (RSS). The RSS consists of only $2^{m+1}$ strategies which are pairwise
uncorrelated or anticorrelated, i.e. the normalized Hamming distance between them is equal to $0.5$
or $1$~\cite{challet98PhysicaA256}. The RSS apparently reduces the complexity and still assures
good quality of prediction as FSS is regularly probed. These statements are visualized in Fig.
\ref{fig:correctness} where marks 'x' and '*' represent results for single agent and all strategies
related to RSS (16 strategies) and FSS (256 strategies) respectively. Despite of a numerous
differences between both pools the results are close, which indicates a superior power of RSS
usage.

The presented above methodology differs significantly from approaches presented by other
researchers~\cite{johnson01PhysicaA299,gou05IEEE,chen08PhysA387}. In those papers the whole bunch
of individuals is used and the strategy distribution is randomly initiated. Although those methods
work either, the approach presented here is optimal.
\subsection{Nonstationary signals and $\lambda$-GCMG predictor}
The agents' behavior is determined by the strategies' utilities which they have at their disposal.
Generally, the strategies that collect more frequently a positive payoff are characterized by a
monotonously rising utilities. The utilities are thus potentially unbounded, what carries some
considerable consequences if the exogenous process is a non-stationary one. Namely, if
characteristics of a modelled object changes at time $t$ then the group of best strategies would
change either. But if, until $t$, some strategies collected many positive payoffs and after time
$t$ they are no longer profitable, then many steps with negative payoffs are required to lower
their utilities. Hence, after time $t$ the system still uses potentially ineffective strategies.
The following example should provide some deeper understanding of this issue.

Let us assume that after 1500 steps the system described by Eq.~\ref{eq:AR3filter} is replaced by
the following one: $y(t) = -0.3y(t-1) - 0.2y(t-2) + 0.6y(t-3) + \xi(t)$.
The new $AR(3)$ process is defined in such a way that the best MG strategies for this process
constitute also the set of the worst strategies of the previous process. Trajectories of utilities
of GCMG predictor are in Fig. \ref{fig:UTrajectory}.
\begin{figure*}[!h]
\begin{tabular}{cc}
\includegraphics[scale=.36]{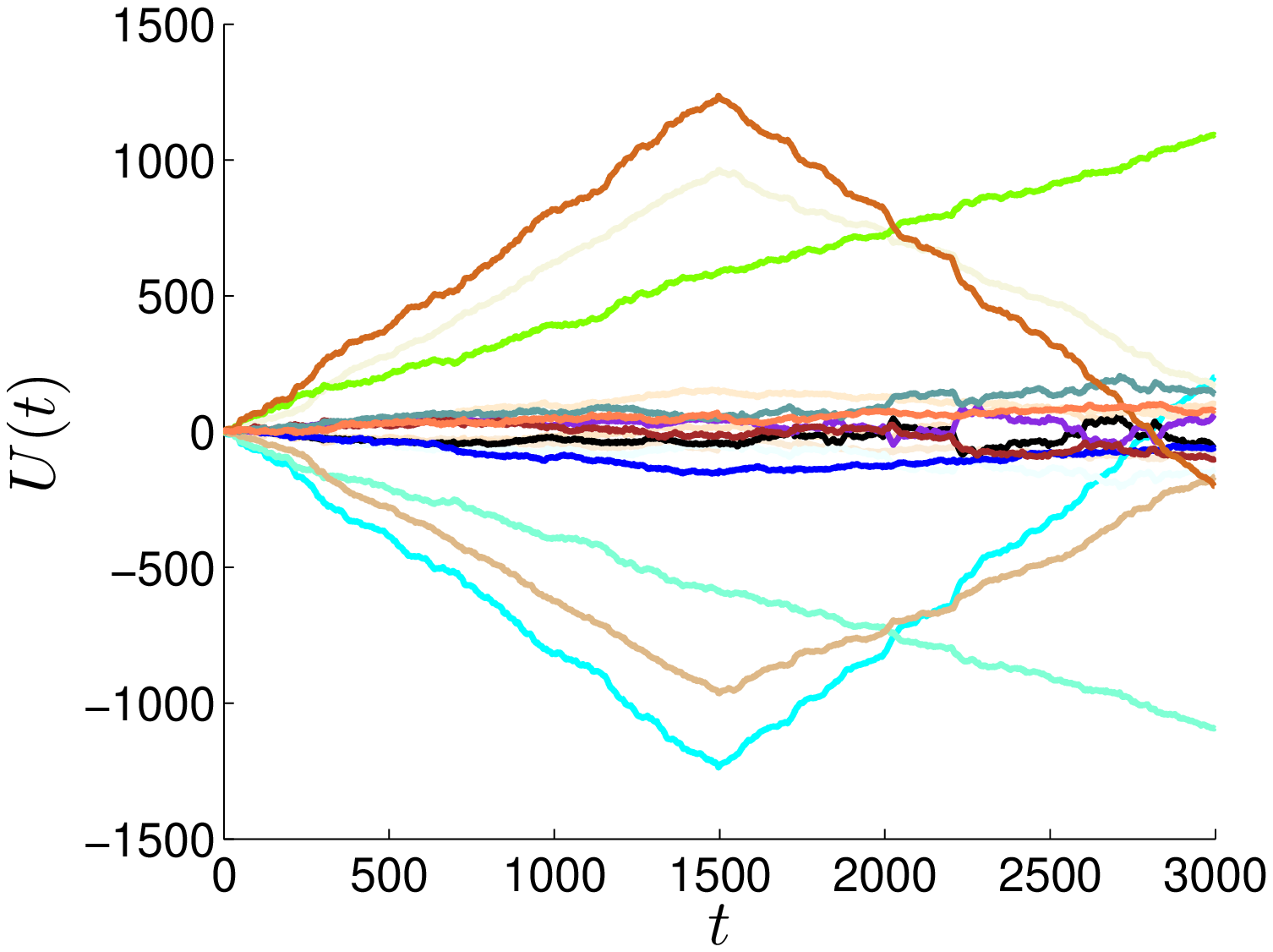} & \hspace{5mm}
\includegraphics[scale=.4]{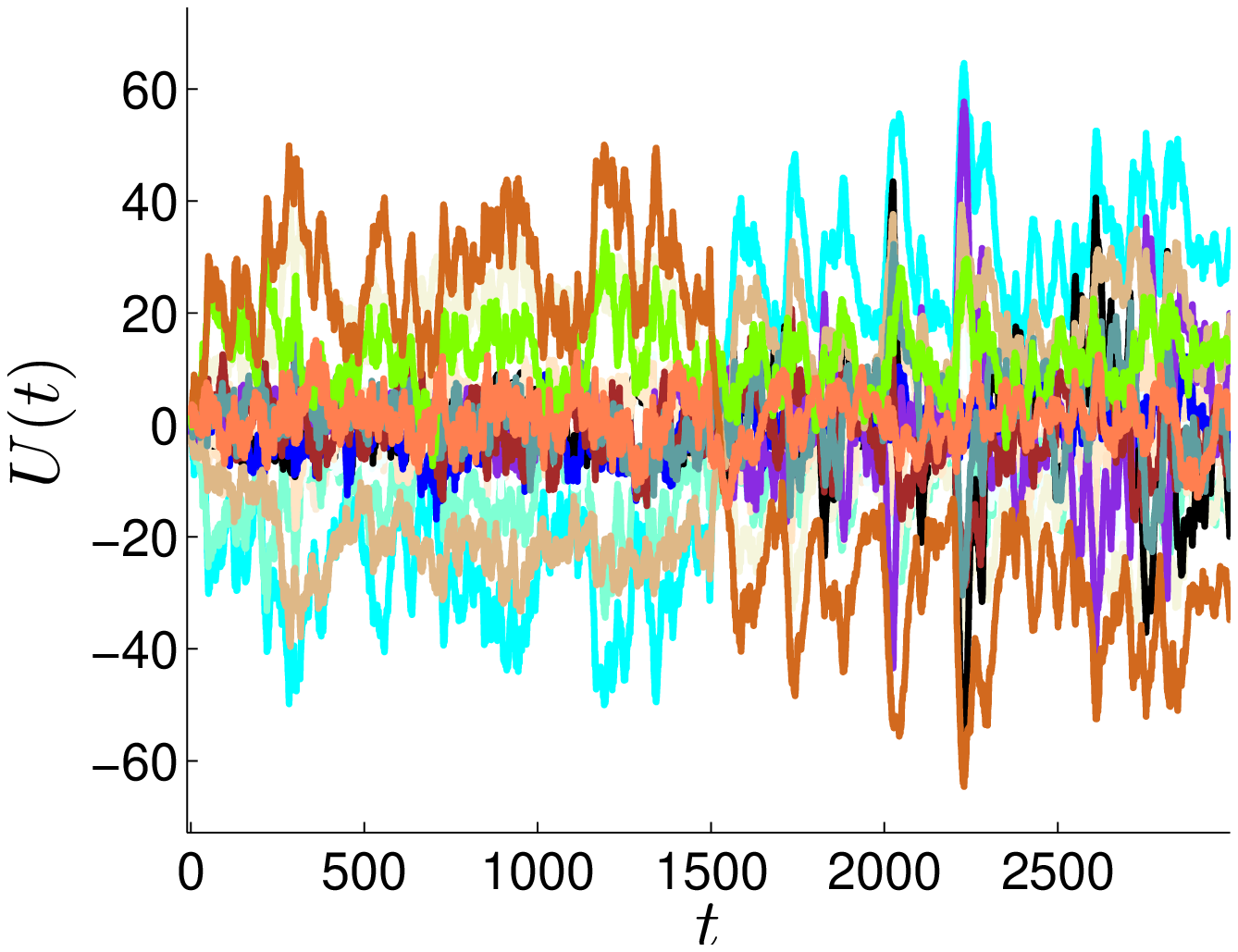}
\end{tabular}
\caption{\em U(t) of all sixteenth pairwise different strategies from RSS for MG characterized by
$N=1$ and $m=3$. The first figure corresponds to a classical GCMG. The second figure corresponds to
the modified $\lambda$-GCMG with $\lambda = 0.97$.} \label{fig:UTrajectory}
\end{figure*}
One can see that after time $t=1500$ the predictor still uses strategy which is no longer
efficient, but is characterized by the highest $U$. As a result, the $\Psi$ as a function of time
starts to decrease, as seen in Fig.\ref{fig:correctnessInTime} (green dotted line).

In order to speed up the new appraisal of outdated $U$ we modified the predictor by introducing
additional parameter $\lambda$ to the rule (\ref{eq:U}) that now is as follows.
\begin{eqnarray}
U_{\alpha_n}(t+1)=\lambda U_{\alpha_n}(t)+\Phi_{\alpha_n}(t), \label{eq:learning_v2}
\end{eqnarray}
where $\lambda \in [0,1]$. If $\lambda=1$ the strategies have an infinite memory of all previous
rewards, what corresponds to standard MG. If $\lambda=0$, then there is no memory effect and the
best strategy is the one with the largest payoff in the previous step. The intermediate values $0<
\lambda < 1$ preserve increasing of $U$ to infinity, when the process is a stationary one, what
considerably speeds up the time of adaptation. The exemplified results for the GCMG with $\lambda =
0.97$ are presented in Fig.~\ref{fig:UTrajectory} (right). All initial conditions are the same as
for picture in the left. It is seen that for $t>1500$ the new best strategy achieves the largest
$U$ much faster than the game without $\lambda$ modification. The system is able to quickly adjust
to changes, and the success rate of prediction is significantly improved, as presented in
Fig.~\ref{fig:correctnessInTime}.

The cost of the introduction of the additional parameter is a need of its optimization. The
heuristic analysis of correctness as a function of $\lambda$ is presented in
Fig.~\ref{fig:correctnessFLambda}. The value $\lambda=0.97$ assures the best results, although
other values, that are close to it, work effectively either.
\begin{figure}[h]
    \begin{minipage}[!t]{0.47\linewidth}
        \centering
            \includegraphics[width=1\textwidth]{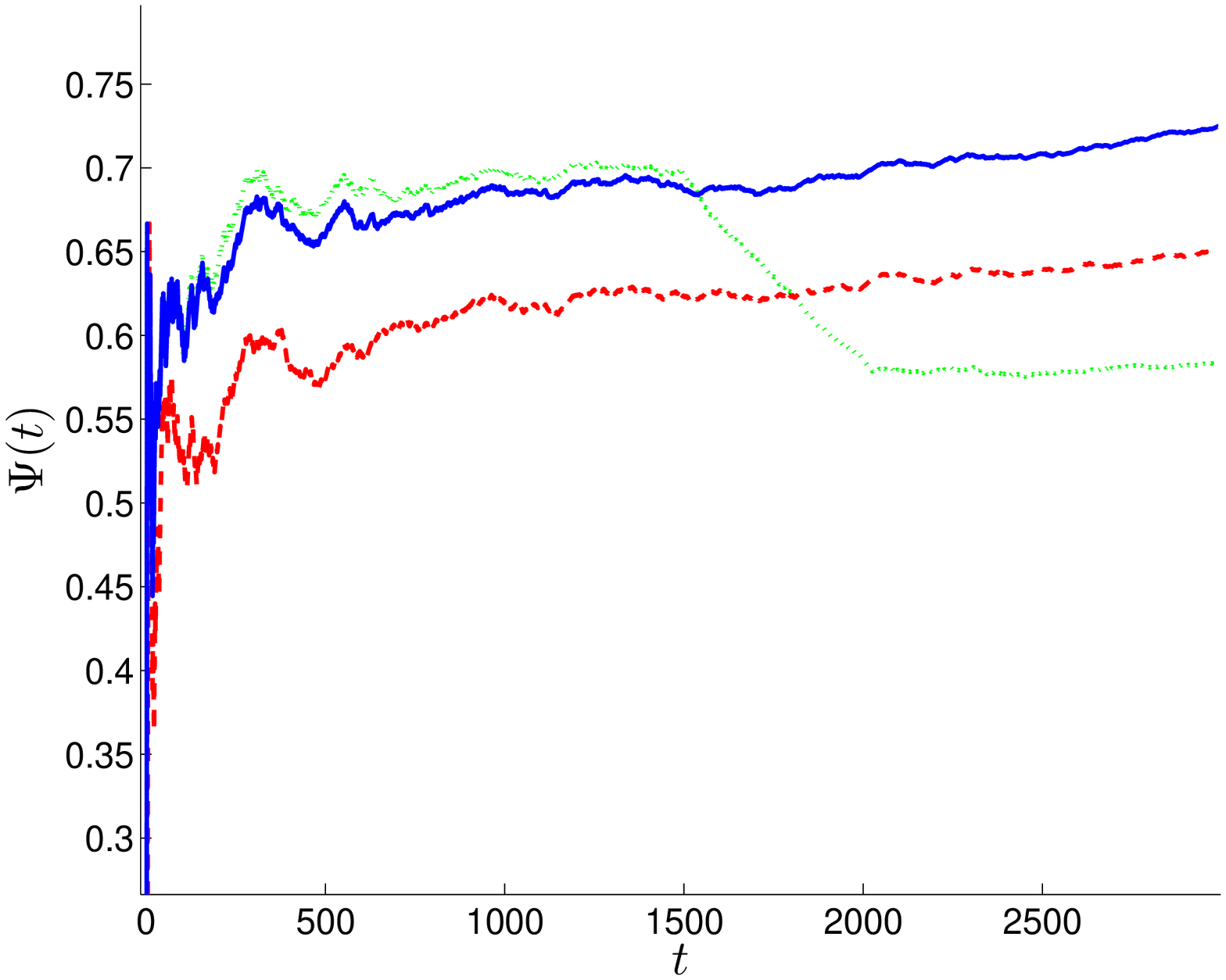}
            \caption{\em The time evolution of $\Psi(t)$ for \mbox{$\lambda$-GCMG} with $N=1$ and $S=16$
            pairwise different strategies from RSS. Red dashed line corresponds to $\lambda = 0.7$,
            solid blue line corresponds to $\lambda = 0.97$ , green dotted line corresponds to $\lambda =
            1$.\newline
            }
            \label{fig:correctnessInTime}
    \end{minipage}
    \hfill
    \begin{minipage}[!t]{0.47\linewidth}
        \centering
            \includegraphics[width=1\textwidth]{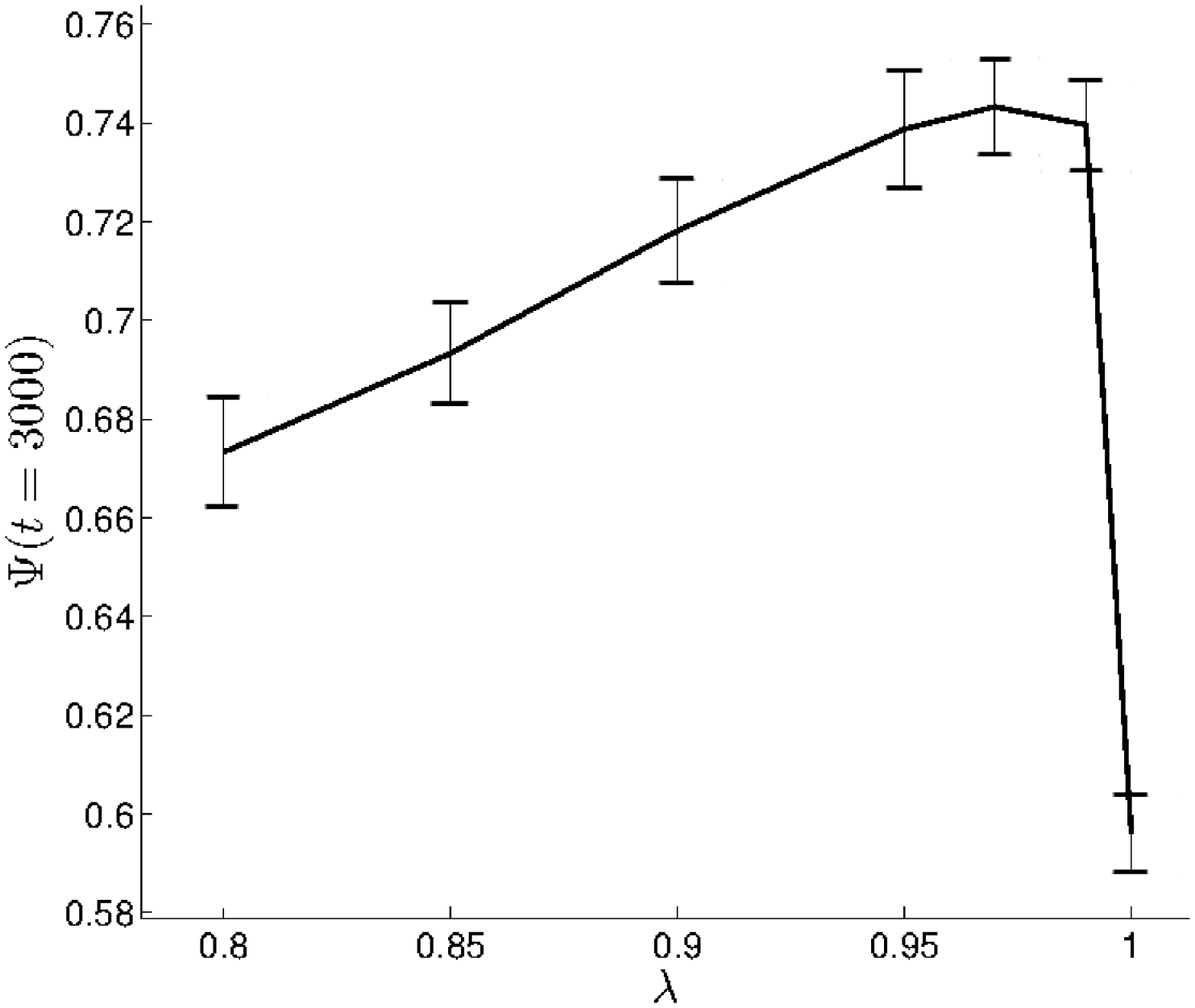}
            \caption{\em The $\Psi(t)$ at time $t=3000$ for \mbox{$\lambda$-GCMG} with $N=1$, $S=16$ pairwise different strategies from RSS and $\lambda = 0.97$. Error bars
            correspond to one standard deviation and curves are drawn to guide ones eye. Statistics are calculated over 10 realizations of nonstationary AR(3) process.}
            \label{fig:correctnessFLambda}
    \end{minipage}
\efig

Summing up, the $\lambda$-GCMG model effectively follows changes in the predicted signal and
significantly outperforms the GCMG for a non-stationary process.
\section{Sign prediction of assets' returns}\label{sec:application}
In this section we present the assumed model of movements of share prices. Given this we apply the
$\lambda$-GCMG predictor to retrieve dependencies between past and future samples of stocks prices
taken from various markets.
\subsection{Price movements model}
We assume that, generally, the dynamics of returns is driven by two factors where one is endogenous
and another one exogenous. The model of return rate is
\be r(t) = f[r(t-1),\ldots, r(t-m)] + \xi(t), \ee
where the first term reflects a dependence on previous returns and the second term represents the
influence of external events.

To be precise, because MG predicts only the sign and not a value of sample, we have to assume even
stronger:
\be \mbox{sgn(r(t))} = f[\mbox{sgn}\ r(t-1),\ldots, \mbox{sgn}\ r(t-m)] + \xi(t).
\label{eq:priceModel}\ee
We assume that the exact form of function $f$ is unknown and it can evolve over time. In case of
the model in Fig.~\ref{fig:blockConfiguration}, the returns $r(t)$ correspond to signals $y(t)$.

Considering the second term $\xi(t)$, there are many various exogenous factors, e.g. publication of
the Gross Domestic Product, inflation, companies annual balances, bankruptcies, etc. Investors
react on them in various ways. Therefore, instead of considering reactions of individuals separately, we
assume that signals of $\xi(t)$ are instances of IID process, with mean value equal to zero. This
signal aggregates reactions of all individuals on all news that appear at time $t$. Such an
approach significantly simplifies further analysis, sice we have to pay an attention only to
dependencies between past and future returns and not to such relations between exogenous factors.

The assumption that exogenous perturbations to returns are IID variables seems more reasonable in
the case of intraday data than for the end-of-day data. As mentioned in the
paper~\cite{johnson01PhysicaA299}, exogenous events are relatively infrequent compared to the
typical transaction rate in the markets. This suggests that the majority in high-frequency data
movements can be potentially self-generated, while the lower frequencies are dominated by exogenous
factors.
\subsection{Application of $\lambda$-GCMG}\label{sec:applicationLambda}
We look for a $\lambda$-GCMG model being able to estimate function $f$ in
Eq.~(\ref{eq:priceModel}). Using our previous results, we set the following parameter values:
$N=1$, $\lambda=0.97$ and the number of strategies covering the RSS. The $m$ value should be chosen
from the range $1-10$. This premise is given by the analysis of data from the London Stock Exchange
(LSE) presented in Refs.~\cite{bouchaud03Fluctuations,lilo04TheLongMemory,gerig07phd}. The precise
choice of $m$ requires some additional analysis. If we take a look at
Fig.~\ref{fig:FW20_correctness_m125}, where the $\lambda$-GCMG is applied as a predictor to FW20
time series i.e. futures contracts on WIG20 index that is the index of the twenty largest companies
on the Warsaw Stock Exchange, then it is seen that better correctness is achieved for lower $m$
i.e. $m=1$.
\begin{figure*}[!h]
\begin{tabular}{cc}
\includegraphics[scale=.28]{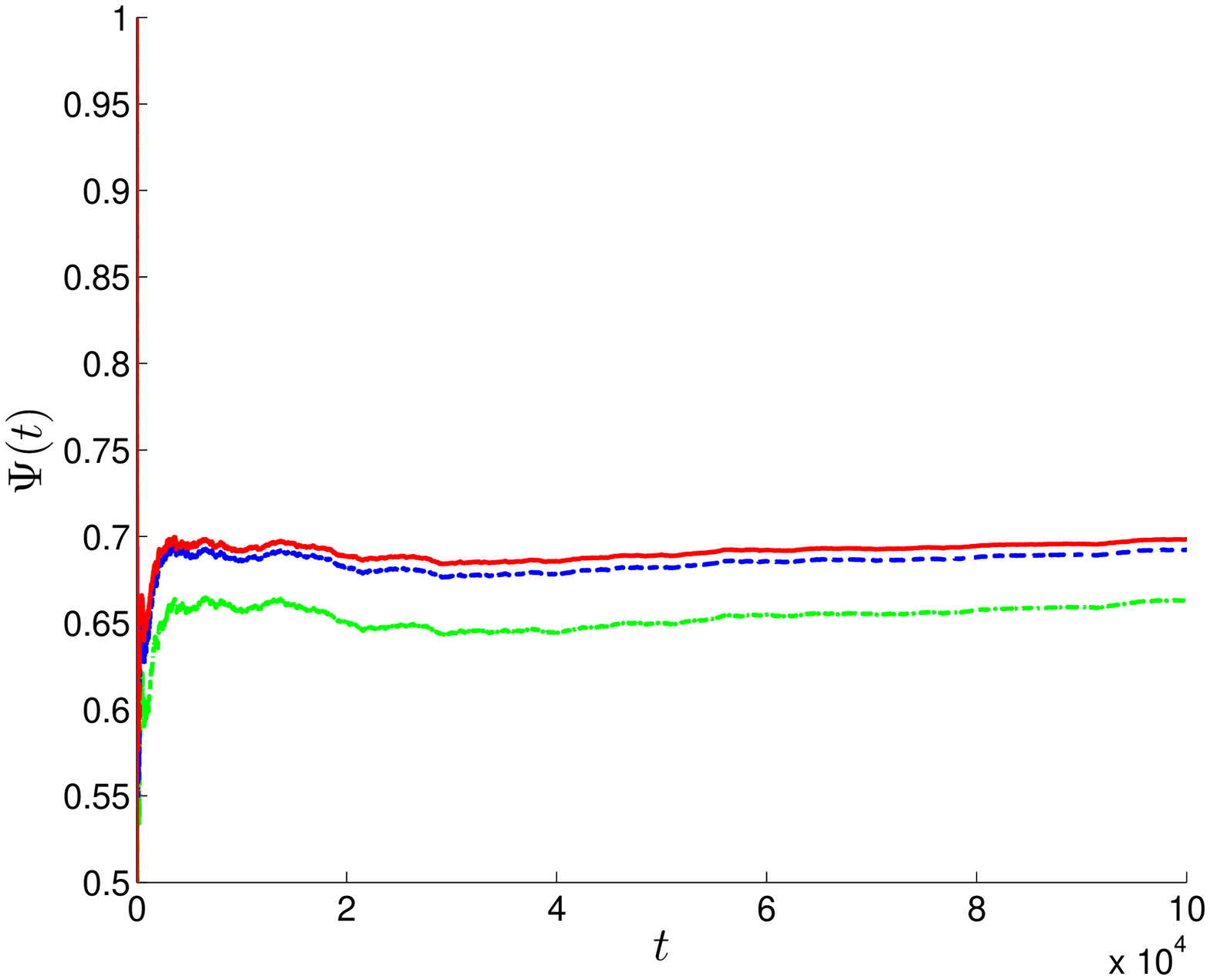}
\includegraphics[scale=.28]{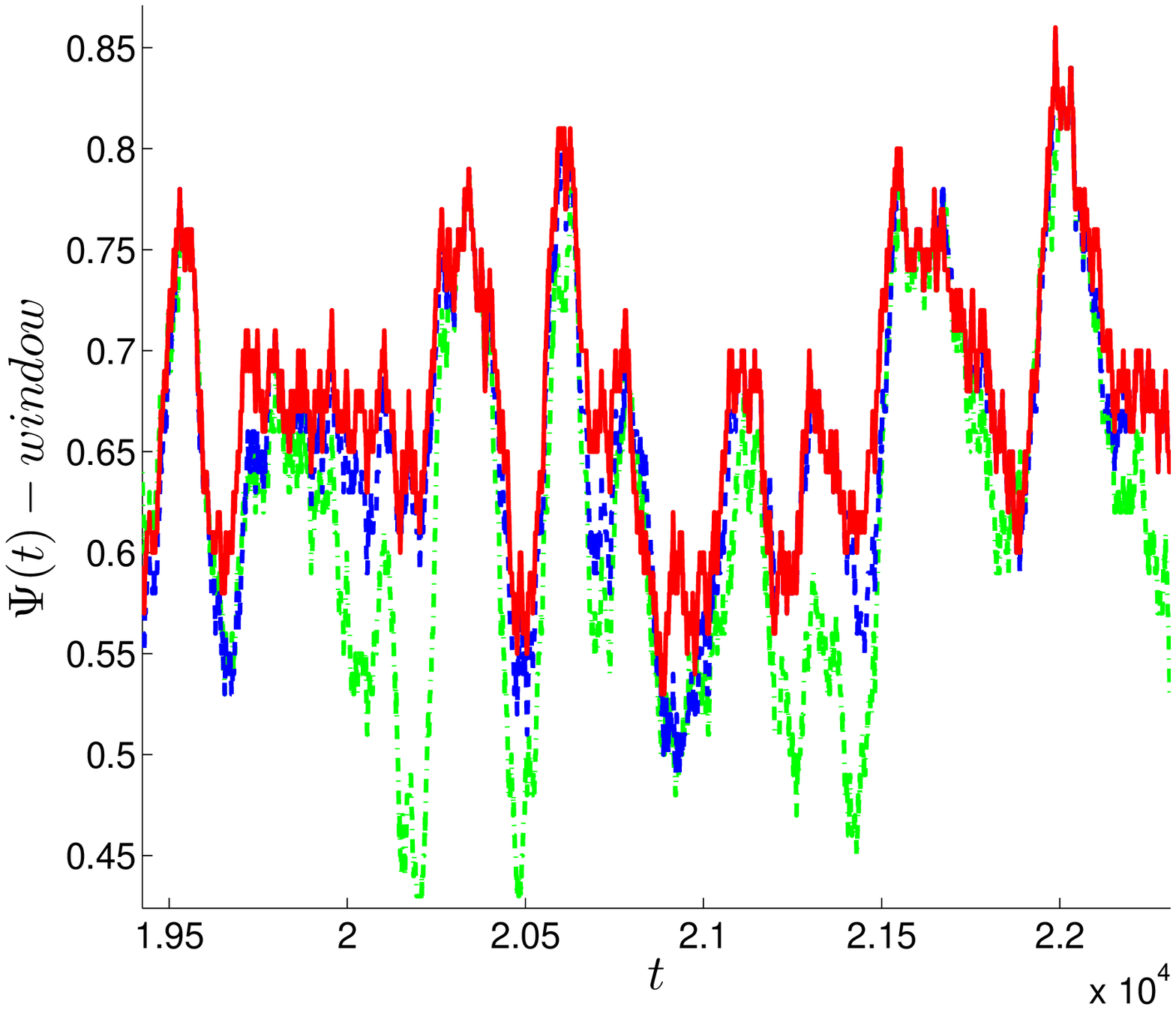}
\end{tabular}
\caption{\em The time evolution of $\Psi(t)$ (left) and $\Psi(t)$ calculated in sliding window of
length equal to $100$ (right) for \mbox{$\lambda$-GCMG} with $N=1$, $\lambda = 0.97$ and strategies
from Reduced Strategy Space. Red solid line corresponds to $m=1$, dashed blue line corresponds to
$m=2$ , green dashed-dotted line corresponds to $m = 5$ (left). The predicted signal is FW20.}
\label{fig:FW20_correctness_m125}\end{figure*}
This result, at least at first sight, seems to be counter-intuitive. The RSS of higher order
includes all strategies of RSS of lower order and additionally introduces the same number of extra
strategies. For example, if $m=2$ then 4 of 8 strategies recognize the same patterns as strategies
for $m=1$ and additional 4 strategies introduce new functionality. The explanation is that for
higher $m$ the additional strategies, although do not capture properly dependencies, get from time
to time higher utility than the basic ones correctly recognizing shorter patterns. This happens
randomly, as sometimes samples are in order that is reflected by some of additional strategies.
However, the pattern does not truly exist. In the next steps the forecast, based on one of
additional strategies, is mostly wrong, what lowers the correctness. In other words, too long
memory spoils the predictor introducing noise of unwanted strategies. The more the memory length
$m$ exceeds real order of the system, the lower the success rate of prediction. The observed
degradation of correctness for longer $m$ indicates that there are no long-range and complicated
patterns in the signal or that their nature is more subtle than MG strategies are able to capture.
The first statement is supported by observations of the autocorrelation of returns, where the only
distinct value is for $\tau = 1$. Similar premise is also included in Ref.~\cite{krause09IDEAL}
where the author found that strategies with shorter $m$ are more preferred by individuals.

The success rate for $m=1$, it is mostly above $0.6$, and from time to time even touches the level
 $0.8$, if calculated in the sliding window. The mean value of correctness is
$0.7$ (see Fig. \ref{fig:FW20_correctness_m125} - left), what seems impressive, at the first sight.
However, we checked that the correctness of Finite Impulse Response (FIR) Wiener
filter~\cite{brown97Intro} of the first order is equal to $0.68$. The use of higher-order filters
does not improve the predictor and, at least in this case, the linear regression is good enough to
assure similar results.

\begin{figure*}[!h]
\begin{tabular}{cc}
\includegraphics[width=0.45\linewidth]{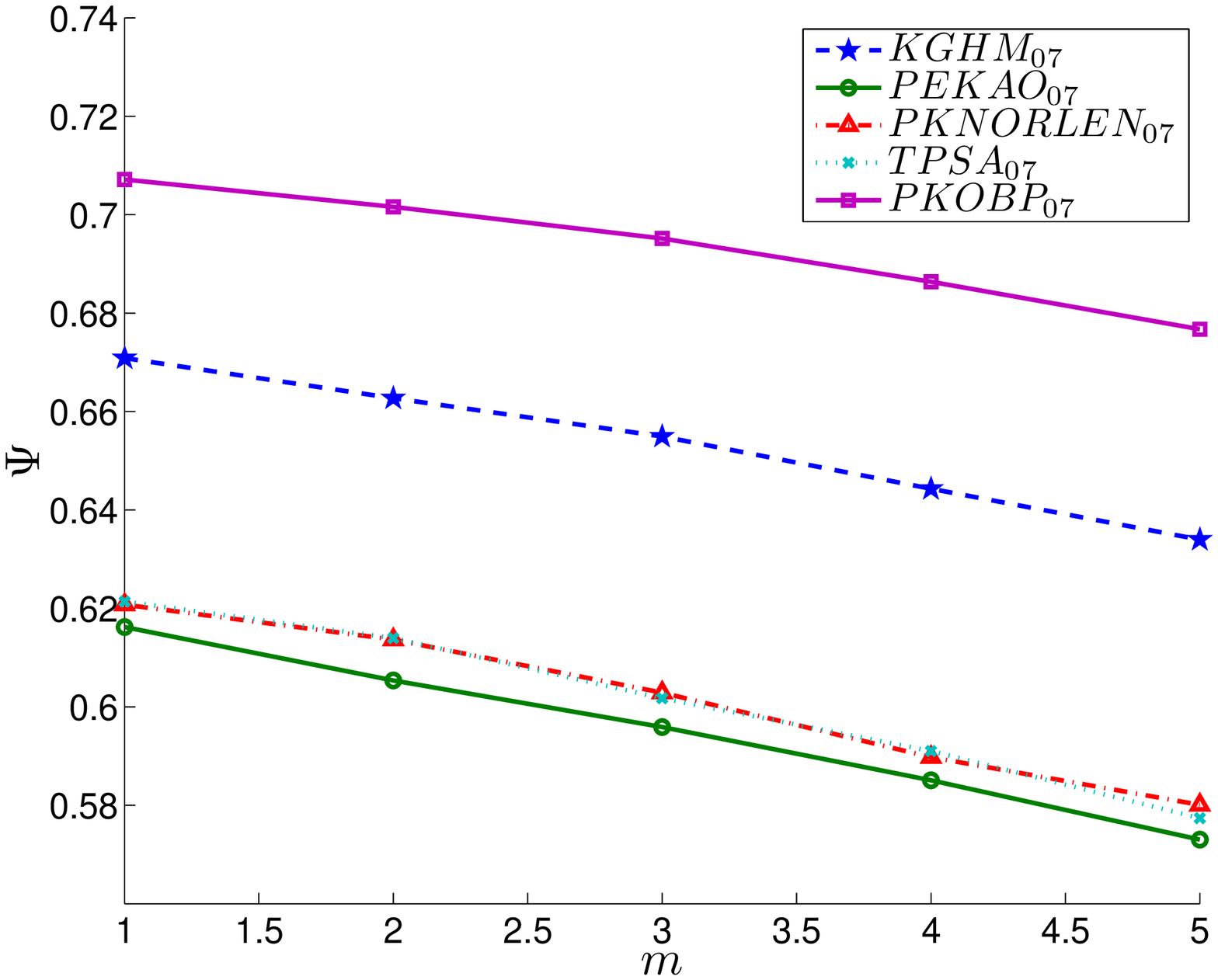} & \hspace{5mm}
\includegraphics[width=0.45\linewidth]{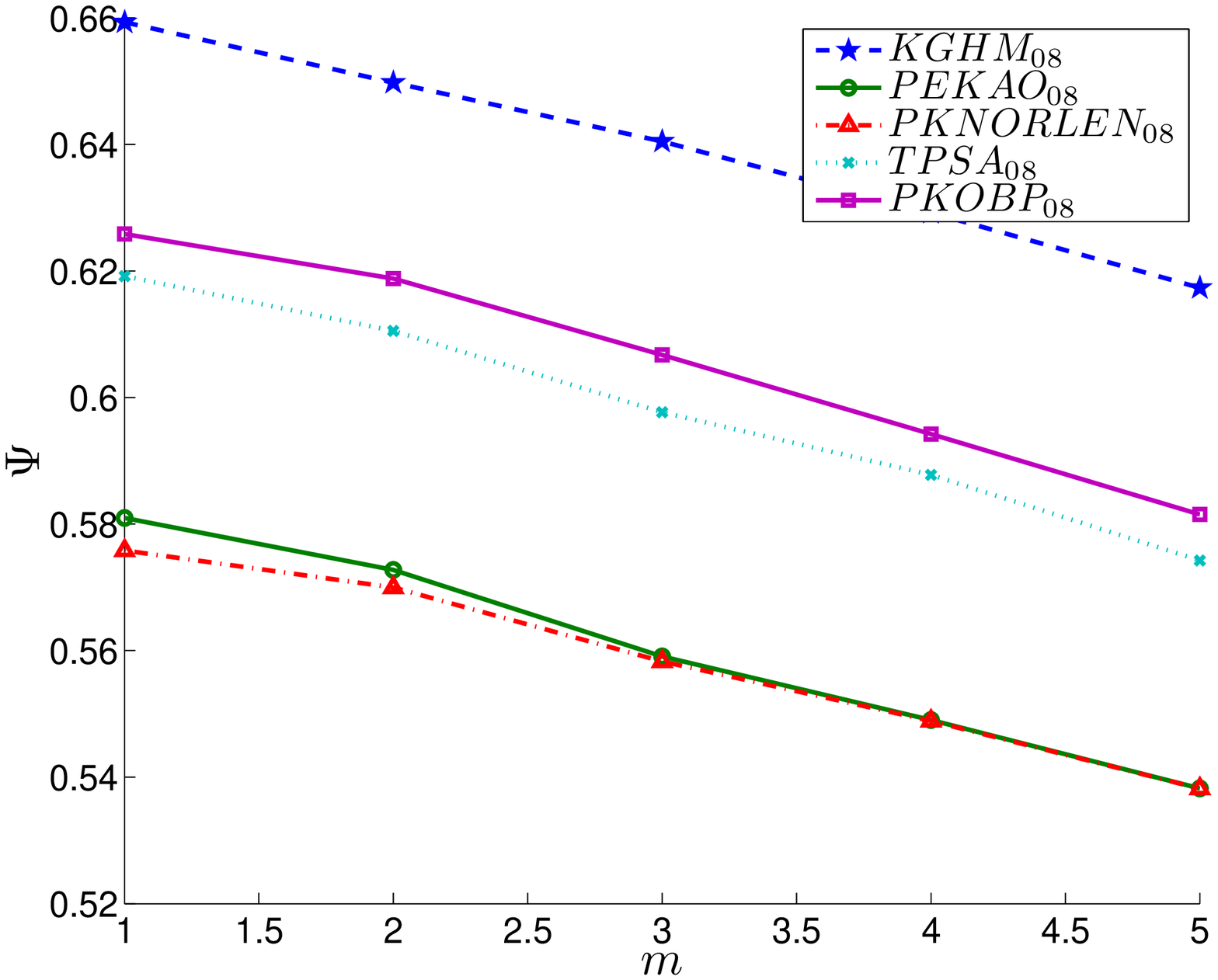}
\end{tabular}
\caption{\em The correctness $\Psi$ as a function of the memory length $m$, for 5 companies with
the biggest impact on index WIG20 for two different years 2007 (left) and 2008 (right).}
\label{fig:correctnessFm}\end{figure*}
The next question is, whether the results achieved for FW20 are specific only for this asset or
they are more universal. We examined separately 5 stocks with the biggest impact on FW20 index:
PKNORLEN, PEKAO, KGHM, PKOBP, TPSA, each of them contributing to the index at the level of
$11-14\%$. As seen in Fig.~\ref{fig:FW20_correctness_m125}, the success rate decreases as a
function of $m$, regardless of the analyzed year. So the results seem to be time and stock
independent.

Another interesting issue is related to the analysis of the best strategy. The question is, whether
there is only one strategy permanently outperforming other strategies or, maybe, different
strategies lead at various moments? If there is only one, it would mean that patterns do not change
over the game or that they do in a way the strategies are unable to capture. One permanently best
strategy also would mean that the extended adaptive version of algorithm is, at least in this case,
as good as an ordinary GCMG or even as good as a linear filter. Indeed, in Fig.
\ref{fig:utilities_m1} (left) it is seen that one of the strategies permanently outperforms others.
Interestingly, this strategy represents the mean-reverting approach, i.e. after history $\mu = 1$
it suggests $a = -1$ and after $\mu = -1$ it suggests $a = 1$. Accordingly, the opposite strategy
representing a trend-follower approach, is the worst one. The results are consistent with
autocorrelation analysis where for $\tau = 1$ the coefficient is negative. We checked that the
results are general for all five examined firms.
\begin{figure*}[!h]
\begin{tabular}{cc}
\includegraphics[width=0.45\linewidth]{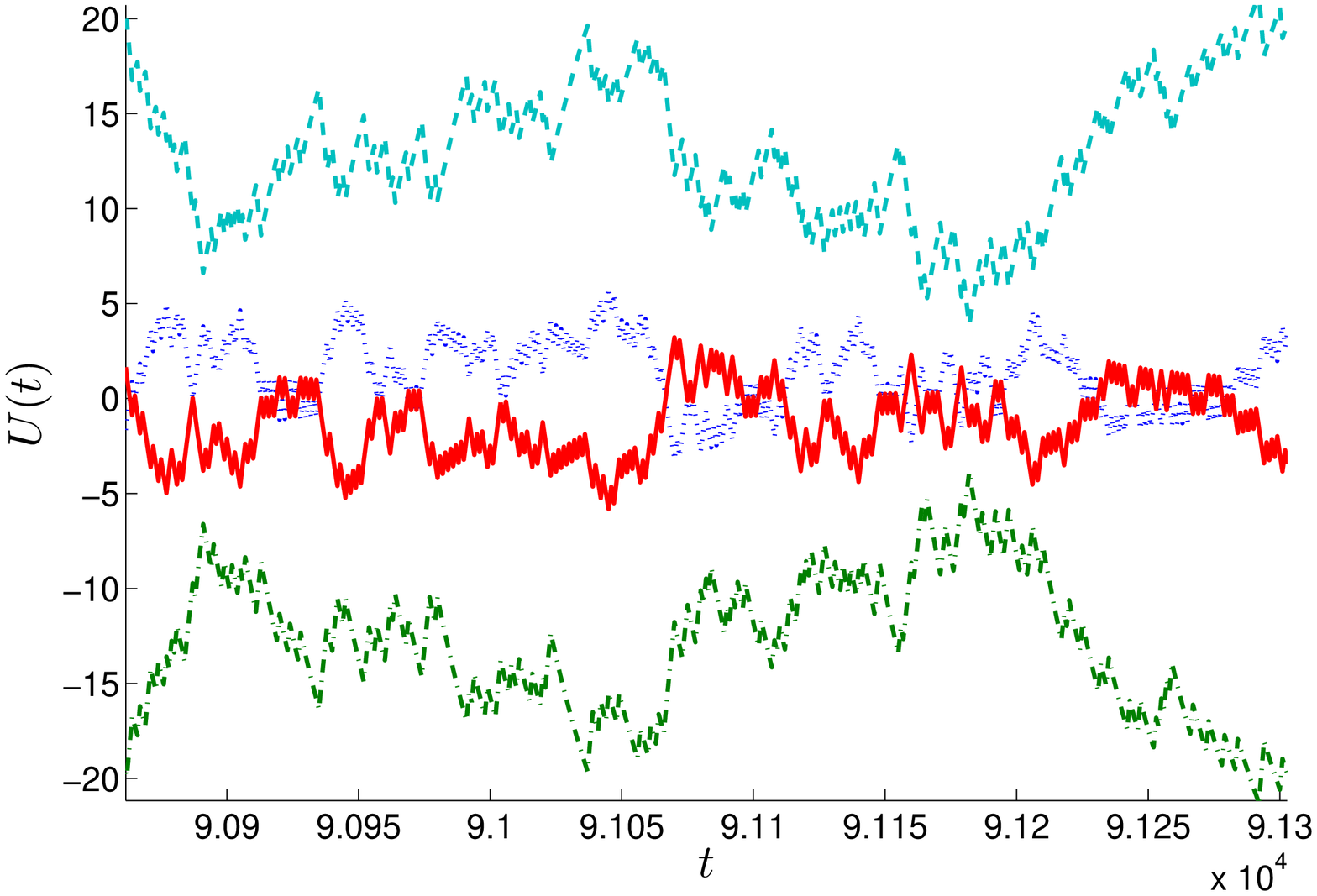} & \hspace{5mm}
\includegraphics[width=0.45\linewidth]{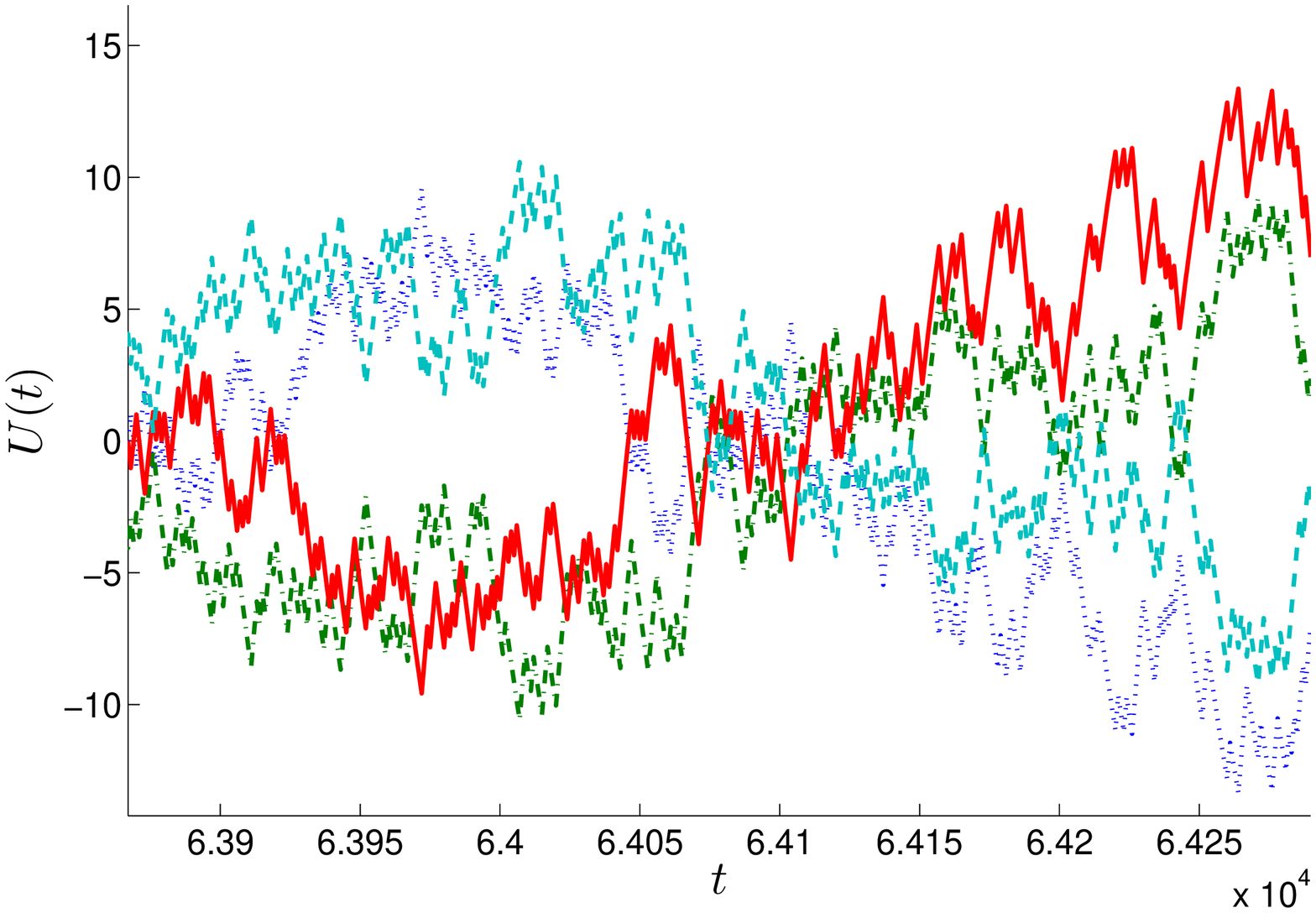}
\end{tabular}
\caption{\em The time evolution of $U(t)$ for all of 4 strategies for \mbox{$\lambda$-GCMG} with
$N=1$ and $\lambda = 0.97$ for index FW20 (left) and Astrazenca company (right).}
\label{fig:utilities_m1}\end{figure*}

The results for Polish market cannot be easily generalized to the London Stock Exchange market. For
example, companies like Vodafone or Astrazeneca are not characterized by only one strategy being
permanently better than others (cf. Fig.~\ref{fig:utilities_m1} - right). The more so, for
Astrazeneca, the best results with the correctness equal to $0.6$ are achieved for $m=2$. It is
difficult to explain why such fundamental differences between stock markets exist. One of the
hypotheses is that the Polish market is less mature than LSE and dependencies between samples are
stronger. Although the hypothesis partially explains the larger values of autocorrelations for
$\tau=1$ in Polish market, it does not explain the longer ranges of the LSE autocorrelations. This
remains an open question.

The above analysis also shows that it is difficult to build a profitable investing system if one
would capture patterns using only agents' strategies. If a sign of a next increment is known to the
investor with encouraging probability, then the investor has to put the order. But placing the
order introduces a perturbation to the forecast. If the order is executed, the system would move to
the next time step and the transaction would be considered as entailing the positive or negative
increment. The investor would predict only its own transactions what, of course, does not assure
any profit. Given this, the prediction for at least two steps is required.
\section{Conclusion}
We applied the minority game as a predictor of an exogenous process. Interestingly, considering
parameters' optimization, we found that the degenerated game with only single agent and all
strategies from FSS is the most efficient configuration. The reason behind is as follows. The FSS
is accessible to every agent and all agents can use all possible patterns. Since there is only one
agent, there is no decision noise from other, badly equipped agents. If using the FSS is
computationally impossible, then the RSS is recommended. Considering the quality of prediction,
minority and majority games are equivalent. Considering non-stationary signals, the parameter
$\lambda$ is introduced in order to speed up the model's convergence. However, the additional
parameter requires extra optimization. This parameter is introduced on heuristic basis.

Applying the predictor to the intraday financial time series allows to effectively perform for one
step where the correctness reaches $70\%$ of properly recognized signs. The best strategy in the
case of Polish stocks is a mean-reverting one. Only slightly worse predictions are attained by
autoregressive systems. Unfortunately, these encouraging results are mostly useless, for building a
profitable investing system. We explained that successful acting requires statistically significant
forecasts for more than only one step forward. The values of autocorrelation function for $\tau
> 1$ suggest that it is hard to achieve. Nevertheless, the complexity of human and algorithmic
actions reflecting price movements consist surely of nonlinear dependencies not properly captured
in the autocorrelation analysis. Therefore, application of MG to the prediction of longer periods
seems to be worth to check. The more so, the idea of building a profitable investing system
requires, in most cases, not only the knowledge of the direction of the price movement but also
some information about the strength of it. Hence, the additional statistical methods should be
compounded with the MG-predictor in order to invest effectively. We consider it as a next step in
our research.

\bibliographystyle{plain}
\bibliography{biblio}

%
%
%
%
%
%
%
\end{document}